\documentclass{article}
\usepackage{arxiv}

\usepackage[utf8]{inputenc} 
\usepackage[T1]{fontenc}    
\usepackage{hyperref}       
\usepackage{url}            
\usepackage{booktabs}       
\usepackage{amsfonts}       
\usepackage{nicefrac}       
\usepackage{microtype}      
\usepackage{lipsum}
\usepackage{graphicx}
\usepackage{float}
\usepackage{cite}
\title{Cost-effective vibration analysis through data-backed pipeline optimisation}

\author{
 Artur Sokolovsky \\
  Department of Design, Manufacturing and Engineering Management\\
  University of Strathclyde\\
  \texttt{artur.sokolovsky@gmail.com} \\
   \And
  David Hare \\
  Bosch Rexroth Ltd.\\
  UK \\
  \texttt{david.hare@boschrexroth.co.uk} \\
  \And
 Jorn Mehnen \\
  Department of Design, Manufacturing and Engineering Management\\
  University of Strathclyde\\
  \texttt{jorn.mehnen@strath.ac.uk} \\

}

\begin{document}
\maketitle
\begin{abstract}Vibration analysis is an active area of research, aimed, among other targets, at an accurate classification of machinery failure modes. This often leads to complex and convoluted signal processing pipeline designs, which are computationally demanding and cannot be deployed in the Edge devices.  
In the current work, we address this issue by proposing a data-driven methodology that allows optimising and justifying the complexity of the signal processing pipelines.
Additionally, aiming to make IoT vibration analysis systems more cost- and computationally effective, on the example of MAFAULDA vibration dataset, we assess the changes in the failure classification performance at low sampling rates as well as short observation time windows.
We find out that a decrease of the sampling rate from 50 kHz to 1 kHz leads to a statistically significant classification performance drop. A statistically significant decrease is also observed for the 0.1 second time windows compared to the 5-second ones. However, the effect sizes are small to medium, suggesting that in certain settings lower sampling rates and shorter observation windows can be used.
The proposed optimisation approach, as well as statistically supported findings of the study, allow a more efficient design of IoT vibration analysis systems, both in terms of complexity and costs, bringing us one step closer to the IoT/Edge-based vibration analysis.
\end{abstract}

\section{Introduction}
Vibration analysis is a hot research topic that is interesting for both researchers and industry practitioners. Successful vibration analysis allows performing machinery condition monitoring, failure mode analysis, and even predictive maintenance. It is especially interesting in the context of the development of Internet of Things systems (IoT), allowing remote system diagnostics. 
However, when developing a practical and scalable vibration analysis IoT solution, one faces certain challenges. Namely, industrial-grade accelerometers can be expensive, making their use in the IoT setting limited. 
Another challenge is the system-specific data properties and failure modes which make generalising the findings across systems non-trivial. 
Finally, there is a general lack of public datasets in the field due to purposeful induction of specific failure modes in a realistic way is challenging and resource-demanding task. 
Due to the limited data availability, there is a demand for a methodology for a data-efficient signal processing pipeline optimisation. It is expected to require little or no extra data for optimisation. Also, it should be robust to overfitting and reasonably fast to evaluate. To the best of our knowledge, no uniform methodology has been proposed so far addressing all these issues. 

The current body of knowledge proposes multiple ways of analysing vibrations and detecting faulty system states~\cite{Antoni2005BlindDemonstrations,Han2019FaultCEEMD,Song2018Vibration-BasedMachinery,Souza2021DeepMachinery,gelman2020novel}. However, many studies are short on demonstrating the optimality of the proposed methods as well as justifying the introduced complexity~\cite{Han2019FaultCEEMD, Song2018Vibration-BasedMachinery,gelman2020novel}. Contributions of every pre-processing step to the final result are often not clear. There are also deep learning-based approaches leading to high classification performance, which are proposed as alternatives to the signal processing methods~\cite{Souza2021DeepMachinery}. They often take the raw data as the model input applying no prior signal processing methods. We believe that it would be beneficial to systematically assess any of these design decisions.
In a recent paper, Rauber et al.~\cite{Rauber2021AnSignals} propose a methodology for machine learning pipeline validation aimed at improving the reliability of the obtained performance.
In the current work, we address a related aspect of the vibration analysis -- the in-depth assessment of the signal processing pipeline components which allows a data-driven justification of the pipeline complexity.
We do so by first optimising individual components of the signal processing stack, then searching for the optimal sequence of the components. This is done using statistical goodness of fit measures. Finally, we demonstrate how machine learning methods can be used in interplay with the proposed approach, getting the best of the two worlds -- machine learning and signal processing.

Using the proposed methodology, we assess the feasibility of the hardware cost optimisation in the industrial IoT setting.
Finally, interested in improving the latency of the classification pipeline, we investigate whether sub-second time windows and linear classifiers (on the example of logistic regression) are suitable for the failure classification of low-mid rotation frequency machinery. 

The contributions of the paper are the following:
\begin{itemize}
    \item We propose a systematic approach to optimisation of the signal processing pipelines on the example of the MAFAULDA dataset~\cite{MAFAULDA};
    \item We assess whether lower sampling rates (1kHz) used by cost effective (and lower sampling rate) accelerometers lead to a statistically significant failure classification performance drop in comparison to the original 50 kHz sampling rate;
    \item Aiming to optimise the latency of the systems, we also investigate whether sub-second observation time windows lead to a statistically significant failure classification performance drop in comparison to the 5 seconds observation time windows;
    \item Lastly, we report the failure classification performance results for logistic regression and CatBoost~\cite{prokhorenkova2017catboost} classifiers, as examples of the simple and robust, and a cutting edge boosting trees estimators, respectively. We compare the performance to the existing studies, as well as allow one to assess in an exploratory way whether simpler and computationally-faster classifiers are suitable for the domain.
\end{itemize}

The rest of the paper is structured as follows: in Section~\ref{sec:methods} we describe the dataset and detail the methodology of the study; in Section~\ref{sec:results} we report the results of the experiments; then we discuss the results in Section~\ref{sec:discussion}; finally, we conclude our work in Section~\ref{sec:conclusion}.
\section{Materials \& Methods}
\label{sec:methods}
The section is structured as follows. We first formulate the aim of the study, then describe the dataset, label design, sampling strategies, and the feature space. Then we list the signal processing methods and the way its hyperparameters and the order are optimised. After the signal processing is done, we apply the machine learning pipeline, including the feature space optimisation, hyperparameter tuning, and model performance evaluation. Finally, we answer the research questions by running statistical tests and computing effect sizes. 

\subsection{Aims}
\label{sec:methods:aims}
In the current study, we introduce the pipeline for systematic optimisation of the signal processing methods and their integration with the machine learning stack.
Using the proposed pipeline, we answer two research questions by formulating the null and alternative hypotheses, then testing the null hypothesis for rejection by running statistical tests. The limitations and generalisation of the findings are discussed in Section~\ref{sec:discussion}. 

\textbf{RQ1:} How does the data sampling rate affect the failure classification performance?
Answering this research question would allow field practitioners to make informed decisions about the choice of the accelerometers and manage the budget more efficiently. 
We propose the following null and alternative hypotheses:
\newline
\newline
H$_{01}$: 50kHz sampled data allows the same or worse classification performance than the 1kHz sampled data. 
\newline
H$_{11}$: 50kHz sampled data allows significantly better classification performance than the 1kHz sampled data. 
\newline
\newline
The lower sampling rate is chosen based on the sampling rates of the cost effective devices available on the market. 

To assess the latency limitations of the vibration analysis pipelines, we propose the second research question. 
\newline\textbf{RQ2}: How does the time window length affect the failure classification performance?
We formulate the hypotheses as follows:
\newline
\newline
H$_{02}$: The failure mode classification performance on the data collected over 5 seconds is the same or worse than for the 0.1 seconds windows. 
\newline
H$_{12}$: The failure mode classification performance on the data collected over 5 seconds is significantly better than for the 0.1 seconds windows.  
\newline
\newline
The reference window size (5 seconds) is the the longest available in the dataset. It is not feasible to study a gradual decrease of the observation window size preserving the optimisation space size. Hence, we set its size to an arbitrary small value, where detection of the system frequencies is still theoretically possible. 

These research questions are valid for the considered system rotation frequencies. The current dataset is considered as a statistical sample of convenience, while the statistical population would involve data of comparable operation frequencies and data quality.  
The model performance is defined as the mean absolute error of the model output on per-class probabilities~\cite{willmott2005advantages}.
\subsection{Dataset}
In the current study, we use the MAFAULDA (MAchinery FAULt DAtabase) vibration dataset, which provides vibration data sampled from two 3-axis accelerometers with a 50kHz frequency~\cite{MAFAULDA}. The data is recorded from a physical system with the capabilities of modelling multiple failure modes, running at frequencies from 11.7Hz to 60Hz and operating on 8-ball bearings.
The dataset is structured as independent 5 second time windows (250k datum entries in each). The time windows are collected at different system rotation frequencies. This gives a detailed picture of how the system operates in different failure modes and a range of rotation frequencies. 

\subsubsection{Label design}
There are 6 high-level system modes of operation: normal, horizontal, and vertical misalignment of the shaft, shaft imbalance, as well as overhang and underhang bearing failures: cage fault, outer race fault, ball fault.
Since this study is not aimed at claiming a state-of-the-art performance in the failure mode classification, but rather demonstrate the approach and answer the research questions, we use the 6 failure modes as the classification labels not distinguishing the failures between different types of bearing failures, but only the corresponding bearing.

\subsubsection{Sampling strategies}
To answer the research questions stated in Section~\ref{sec:methods:aims} of the paper, we sample the raw accelerometer readings in three ways: 
\begin{enumerate}
    \item To mimic the real-world limitations of the cost effective IoT vibration analysis solutions, we sample every 50th data point, representing the 1kHz sampling frequency. The full available observation time window is used, leading to 5k data points entries.
    \item Aiming to study how the window length affects the performance, we obtain the first 5k points with a 50kHz sampling rate, corresponding to the 0.1 seconds observation time window.
    \item First and second sampling strategies are compared to the full available time window of 5 seconds and the maximum available sampling rate of 50kHz.
\end{enumerate}

\subsubsection{Feature space}
The existing body of knowledge proposes various feature spaces for vibration analysis, including using raw vibration readings in deep learning models and designing domain-specific feature spaces. In the current work, we aim to make the findings of the study generalisable, hence we use an open-source Python-based software package for time series feature extraction -- TSFEL~\cite{barandas2020tsfel}, version 0.1.4 (the latest available at the moment of conducting the experiments). We use its spectral-domain features. 
This feature space is used uniformly at all the stages of the study -- in the signal processing and machine learning pipelines.

\subsubsection{Cross-validation}
When reporting the results, we split the data into training and test batches using a K-fold approach with 3 folds. The data is split into training and test 3 times so that the test parts of the folds reconstitute the whole dataset. This allows us to get test results for the whole dataset. We split the data into folds before any optimisations and preserve the splits across all the stages. The predictions are concatenated across the folds before computing the effect sizes and running the statistical tests.
Machine learning optimisation is done on the training batch and at every step includes internal cross-validation policy with the 3-fold split. We withhold from running the nested cross-validation as this would significantly increase the computation times making the experiments infeasible. 

\subsection{Signal processing pipeline}
Guided by the existing body of knowledge, we take the components of a signal processing pipeline proposed for lower frequency systems~\cite{Han2019FaultCEEMD} and comprised of continuous wavelets~\cite{rioul1992fast}, non-linear energy operator~\cite{kaiser1993some}, CEEMD~\cite{torres2011complete}, and envelope transformation~\cite{feldman2011hilbert}. 
The study does not systematically optimise the hyperparameters and the order of the steps. In our design, we address this point by assessing the performance of the individual pipeline components and different component sequences. Since CEEMD is computationally very demanding, we use its simpler version -- EMD~\cite{huang1998emd}. 

EMD is an adaptive noise reduction approach that is commonly applied to non-linear and non-stationary time series. It breaks the non-stationary time series into intrinsic, finite components or Intrinsic Mode Functions (IMFs).
The non-linear energy operator computes the energy of the signal in a non-linear fashion, allowing it to successfully capture the characteristics of the systems. Being non-linear, it allows spotting properties of the data missed by the Fourier transform.
Wavelets are a widely used method in many fields, starting from data compression, ending with image processing. They are also commonly used for data approximation and denoising, where the unwanted frequencies are filtered out.  

\begin{figure}[H]
    \includegraphics[width=0.7\textwidth]{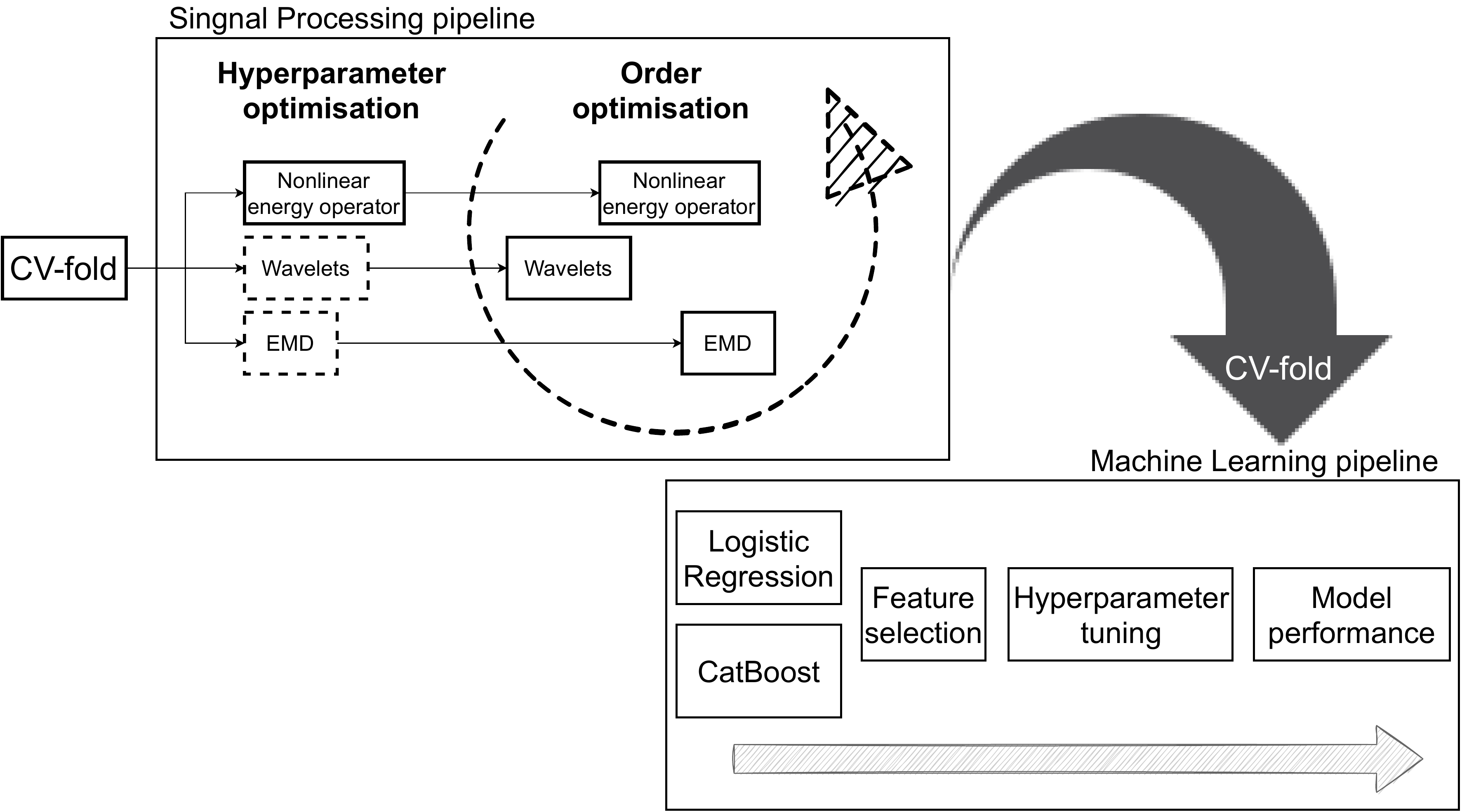}
    \caption{Diagram illustrates the optimisation steps of the signal processing and then machine learning (ML) pipelines. The dashed lines in the signal processing part represent elements optimised at the considered stage. ML pipeline is run from left to right sequentially. After the optimal configuration for a signal processing pipeline is found, the dataset is processed and fed into the ML pipeline. At every step, a fixed set of features is extracted from the data and logistic regression is fitted and its goodness of fit is assessed (Akaike's Information Criterion). The CV-fold corresponds to the cross-validation folds used throughout the study.}
    \label{fig:dia}
\end{figure}

Since the whole process is computationally intensive, we make a design decision to exclude envelope transformation from the list of the methods. Consequently, we end up with the signal processing pipeline comprising of 3 methods to optimise: wavelets, non-linear energy operator, and EMD. In the pipeline, we do not make any assumptions about the optimal sequence of the methods, instead, we search for it by iterating over all possible variants. In the case of the larger component spaces, it becomes infeasible and some constraints should be introduced. 
The high-level diagram of the steps is provided in Figure~\ref{fig:dia}.

Since the detection of failures can be done by manually assessing the frequency domain of the vibration spectra and associating particular peaks with the failures, we assume that linear effects are strong in the data. Hence, we use a logistic regression model and Akaike's Information Criterion (AIC)~\cite{sakamoto1986akaike} to assess its goodness of fit for the signal processing optimisations. AIC allows us to compare across different signal processing pipeline configurations -- both hyperparameters and the order of the methods. 
While one could use more advanced methods involving cross-validation, it would have been more computationally intensive and also requiring extra data. Considering the setting and the complexity of getting high-quality labelled data, we aim to minimise the use of extra data in the optimisation process.

\subsubsection{Optimisation}
As the first step, we optimise the hyperparameters of the pipeline components. We discuss the weaknesses of this design decision in the \nameref{sec:discussion:limitations} section.
Namely, there are two which require parameter optimisation: wavelets and EMD. Wavelets are optimised by the scale range and the type of wavelet, and EMD -- by the number of intrinsic mode functions (IMFs). 

After the hyperparameters are tweaked, there is a need to justify the order of the applied methods. To do so, we compute the goodness of fit on all the possible permutations of the methods, including the goodness of fit on the feature space without any signal processing applied.
For both stages we remove the 0-variance feature and standardise feature values before feeding into the logistic regression.

\subsection{Machine learning pipeline}

After the signal processing pipeline is optimised and its complexity is backed by the data, machine learning methods are applied to the pre-processed data. 

When performing the experiments, we want to cover the performance-oriented and latency-oriented scenarios. Hence, we classify the data using logistic regression and CatBoost classifiers~\cite{prokhorenkova2017catboost}. 

\subsubsection{Feature selection}

While we do not perform any feature space optimisation at the point of signal processing, there is a need to do so if one aims to get good classification performance. We use recursive feature elimination with cross-validation (RFECV) with a decreasing step and class-weighted f1-score as the performance metric~\cite{guyon2002gene}. While there are less computationally intensive methods, RFECV is widely accepted across the ML community. It also aligns well with the data-backed approach of the current study. 

\subsubsection{Hyperparameter tuning}

After the feature space is optimised, there is a need to adjust the hyperparameters of the models. In the logistic regression, we optimise the regularisation parameter, and for CatBoost -- the maximum depth of the trees, regularisation, and the number of trees.

\subsubsection{Model evaluation}

To give a complete picture of the model performance, we report multiple classification performance metrics. Namely, we report f1-score, precision, recall -- weighted by class, and accuracy. F1-score is known to work well with class-imbalanced datasets, precision and recall are reported as its components~\cite{aurelio2019learning}. 
Accuracy provides the fraction of correctly classified entries. We claim that accuracy is a valid metric to use in the considered case as the class imbalance in the dataset is moderate (reported in Table~\ref{tab:dataset}) and the majority class will not lead to the artificially high metric values.  

\subsection{Statistical evaluation}

We perform the analysis on the absolute differences between the predicted class probabilities and the actual labels in a per-entry fashion. Hence, for each classified entry, we obtain a vector of positive values from 0 to 1, of the length equal to the number of classes. From there we compute the mean absolute error (MAE) for the whole dataset.
To formally answer the research questions, we perform hypothesis testing -- this allows us to formally reject the null hypotheses. Also, we compute the effect sizes to quantify the strength of the effect. Moreover, by reporting the .95 Confidence Intervals (CIs), we generalise the effect sizes to the statistical population. 

As an effect size measure, we use Hedge's $g_{av}$. It computes the mean group difference corrected for paired entries~\cite{lakens2013calculating}. 
We test the hypotheses using the Wilcoxon test~\cite{wilcoxon}, which is a non-parametric version of a t-test, which is applied in cases of paired measurements and different variance between the studied groups.

Since we report multiple statistical tests on related data, it is necessary to account for the multiple comparisons. We do so by correcting the significance levels of the tests using Bonferroni corrections~\cite{bonferroni1936teoria}. Since all the tests are performed on the related data, we treat all the tests of the study as a single experiment family. Hence, the corrected significance level is $\alpha_{corr}=\alpha/n$, where $n$ is the number of tests. Corrections for multiple comparisons are also applied to the confidence intervals of the effect sizes. 

\section{Results}
\label{sec:results}
\subsection{Dataset}
The distribution of the failure modes by the number of collected entries is provided in Table~\ref{tab:dataset}. As it can be seen, around half of the entries are related to the bearing failures and the least number of entries is available for the normal machine state.

\begin{table}[H] 
\caption{The table provides details on the dataset, as it was used in the study. \label{tab:dataset}}
\begin{tabular}{c|c}
\toprule
\textbf{Failure mode}	& \textbf{Number of entries}\\
\midrule
Normal		& 49 \\
Horizontal misalignment (shaft)		& 197 \\
Vertical misalignment (shaft)		& 301  \\
Shaft Imbalance & 333 \\
Overhang bearing & 513 \\
Underhang bearing & 558 \\
\midrule
\textbf{Total} & 1951\\

\bottomrule
\end{tabular}
\end{table}

\subsection{Signal processing pipeline}
The optimisation of the signal processing methods is done for two sets of hyperparameters, depending on the number of data points (Table~\ref{tab:sigProcessParams}). Namely, for the 250k data points, we shrink the number of hyperparameter values to stay in the feasible range of required computation power. The implications of this design decision are discussed in the limitations of the paper. Since the experiments are parallelisable across the data entries, we use a High Performance Computing cluster to run the experiments, taking around 20k CPU hours in total. 

\begin{table}[H] 
\caption{The table provides details of the signal processing hyperparameter optimisation for EMD and wavelet methods. The parameters in the square brackets correspond to the sets of values used in the search space. 
The -1 value corresponds to second largest available value and None indicates an absence of the bound. The 'shrunken' space is used for the 250k data points configuration and 'full' -- for the 5k points. \label{tab:sigProcessParams}}
\begin{tabular}{c|c|c}
\toprule
\textbf{Method}	& \multicolumn{2}{c}{\textbf{Parameter values}}\\
{} & Shrunken & Full \\
\midrule
EMD, IMF lower bound & [0,1] & [0,1,2,3,4,5] \\
EMD, IMF upper bound & [-1, None] & [6,7,8,9,None] \\
Wavelet, lower scale & $2^{[0,1]}$ & $2^{[0,1,2,3]}$ \\
Wavelet, upper scale & $2^{[9]}$ & $2^{[5,7,9]}$ \\
Wavelet, type & [Morlet, Gaussian] & [Morlet, Gaussian] \\
\bottomrule
\end{tabular}
\end{table}

The resulting hyperparameters of EMD and wavelets are provided in Table~\ref{tab:sigProcHyperOptRes}. One sees that EMD has different parameters for 5 second and 0.1-second windows -- there is an upper bound in the IMFs in 2 of 3 cases of the 0.1-second window and no such bound is observed for the 5-second observation window experiments. Wavelets filter out the first scale for the 50kHz sampling rate, 0.1s window in 2 of 3 folds, as well as in all the folds of 1kHz, 5s window. The first wavelet scale corresponds to the highest frequencies out of the spectrum.

\begin{table}[H] 
\caption{The table shows the final EMD and wavelet hyperparameters for all the experiments and cross-validation (CV) folds. The parameters are reported as "IMF lower bound, IMF upper bound" for EMD and "lower scale, upper scale, type" for wavelets.  \label{tab:sigProcHyperOptRes}}

\begin{tabular}{c|c|c|c}
\toprule
{\textbf{CV fold}} & \multicolumn{3}{c}{\textbf{EMD}}\\
{} & 50kHz, 0.1s & 50kHz, 5s & 1kHz, 5s \\
\midrule
0 &      0, None &         0, None &        0, None \\
1 &         0, 8 &         0, None &        0, None \\
2 &         0, 9 &         0, None &        0, None \\
\bottomrule
{} & \multicolumn{3}{c}{\textbf{Wavelets}}\\
\midrule
0 &   $2^1$, $2^9$, Morl &      $2^0$, $2^9$, Morl &     $2^1$, $2^9$, Morl \\
1 &   $2^1$,$2^9$, Morl &      $2^0$, $2^9$, Morl &     $2^1$, $2^9$, Morl \\
2 &   $2^0$, $2^9$, Morl &      $2^0$, $2^9$, Gaus &    $2^1$, $2^9$, Morl \\
\bottomrule
\end{tabular}
\end{table}

In the second stage of the signal processing optimisation we aim to find the optimal order of the applied methods. The results are provided in Table~\ref{tab:sigProcOrderRes}. In the optimised configuration, both EMD and wavelets are applied consistently across all the folds of the 0.1s window experiments. The results are less consistent for the other two configurations: no pre-processing is done in 2 of 3 folds for 5s window, 50kHz sampling rate, and in 1 fold of 1kHz, 5s window. Finally, the non-linear energy operator was not applied to any of the folds of any of the datasets.

\begin{table}[H] 
\caption{The table shows the final order of the signal processing methods applied for all the experiments and cross-validation (CV) folds. The sequences are reported in the same order they are applied. 'None' corresponds to no signal processing performed. \label{tab:sigProcOrderRes}}
\begin{tabular}{c|c|c|c}
\toprule
\textbf{CV fold} & \textbf{50kHz, 0.1s} & \textbf{50kHz, 5s} & \textbf{1kHz, 5s} \\
\midrule
0 &  EMD, wavelets &     EMD, wavelets &              EMD \\
1 &  EMD, wavelets &              None &             None \\
2 &  EMD, wavelets &              None &              EMD \\
\bottomrule
\end{tabular}
\end{table}

\subsection{Machine learning pipeline}
When optimising the machine learning pipeline, we follow a uniform procedure for both estimators and all the experiment configurations. 
The feature space dimensionality obtained from the TSFEL package is 2016 input features and is consistent across all the experiments.
Due to the large feature space, the feature selection is done in an annealing fashion, where the RFECV step decreases with the decrease of the feature space size. The feature number thresholds are: 700, 350, 125, 75, 37, 17, 8 and the corresponding steps are: 400, 100, 50, 25, 12, 6, 3.
The final sizes of the feature space for both estimators, all the cross-validation folds, and experiments are provided in Table~\ref{tab:RFECVresult}. Being a linear estimator, logistic regression on average is optimised to smaller feature spaces. Being an ensemble estimator, constructed of unstable base estimators (decision trees), CatBoost has more diversity in the feature space between folds.

\begin{table}[H] 
\caption{In the current table we report the final numbers of features after the feature selection process for the 3 cross-validation folds, CatBoost and Logistic Regression estimators and all the experiments. The experiments are encoded as "sampling frequency, time window length". \label{tab:RFECVresult}}
\begin{tabular}{c|c|c|c}
\toprule
\textbf{Estimator}	& \multicolumn{3}{c}{\textbf{Experiment}}\\
{} & 50kHz, 5s & 50kHz, 0.1s & 1kHz, 5s \\
\midrule
CatBoost & 16, 268, 716 & 191, 2016, 666  & 241, 566, 164  \\
Logistic Regression & 125, 200, 150  & 75, 200, 122  & 200, 263, 350  \\

\bottomrule
\end{tabular}
\end{table}

The estimator hyperparameters are optimised after the feature selection. The particular optimised hyperparameter values are available in the published reproducibility package and not reported in the main body of the manuscript as largely influenced by the optimised feature spaces and can hardly be directly related to the results of the study. 
The classification performance for both estimators is provided in Table~\ref{tab:classifResult}. At this point, there is no observable performance supremacy in any of the configurations. The worst performance is observed for 50kHz sampling rate, 0.1s window, and the other two configurations perform similarly.  

\begin{table}[H] 
\caption{In the current table we report the f1-score, precision, recall and accuracy (Acc.) of the estimators for all the experiments. F1-score, precision, recall are weighted by class. \label{tab:classifResult}}
\begin{tabular}{c|cccc|cccc}
\toprule
{\textbf{Experiment}} & \multicolumn{4}{c|}{\textbf{Logistic Regression}} & \multicolumn{4}{c}{\textbf{CatBoost}} \\
{} &   F1 &  Precision &  Recall &  Acc. &   F1 &  Precision &  Recall &  Acc. \\
\midrule
1kHz, 5s  & 0.99 &       0.99 &    0.99 &      0.99 & 1.00 &       1.00 &    1.00 &      1.00 \\
50kHz, 5s & 1.00 &       1.00 &    1.00 &      1.00 & 0.99 &       0.99 &    0.99 &      0.99 \\
50kHz, 0.1s    & 0.94 &       0.97 &    0.93 &      0.93 & 0.91 &       0.94 &    0.90 &      0.90 \\
\bottomrule
\end{tabular}
\end{table}

\begin{figure}[H]	
\includegraphics[width=0.7\textwidth]{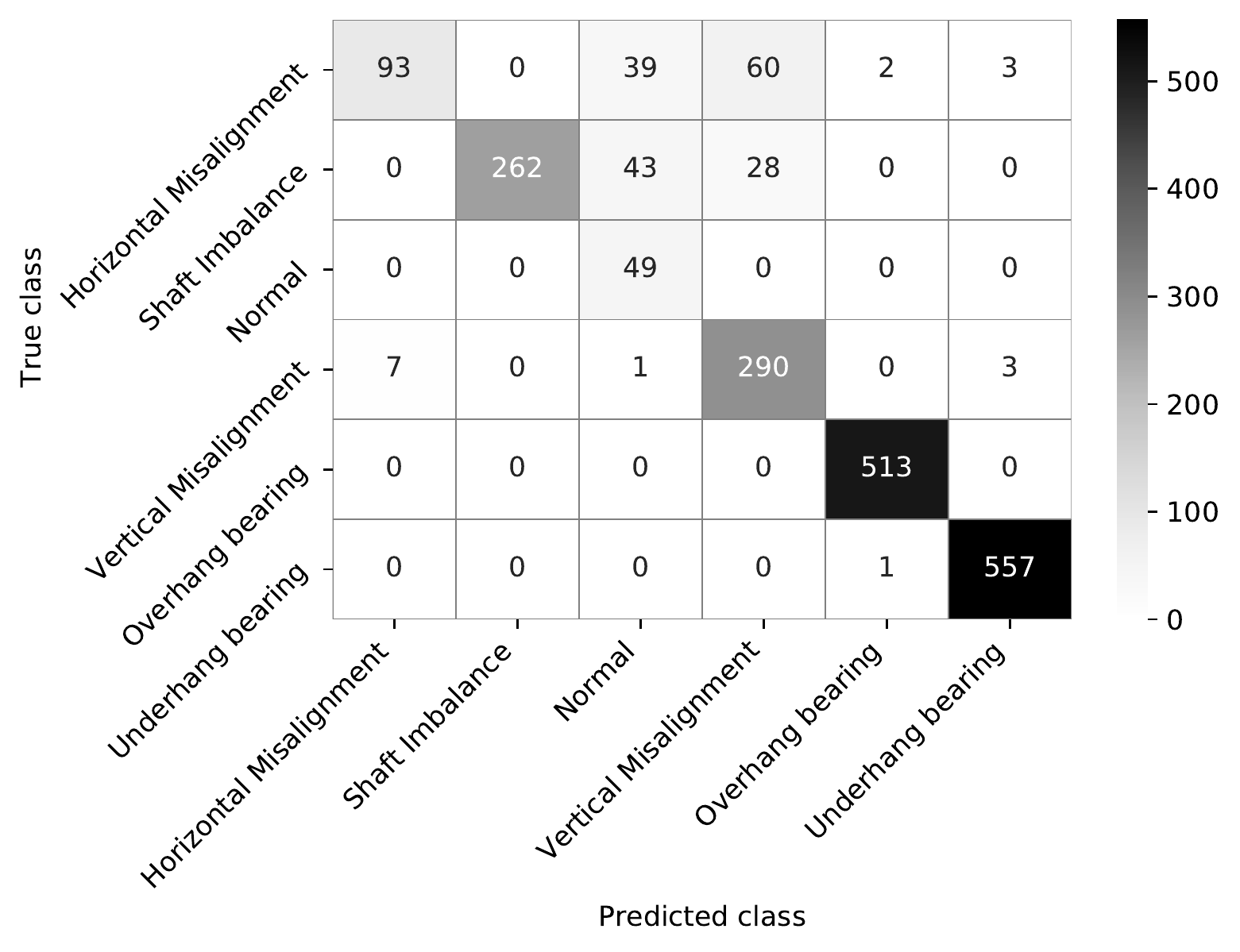}
\caption{We present the confusion matrix of the worst performing experiment -- 50kHz sampling rate, 0.1s time window, CatBoost classifier. \label{fig:confMat}}
\end{figure}

In order to investigate the classification patterns, we report a confusion matrix of the worst-performing experiment configuration -- 50kHz sampling rate, 0.1 second time window, CatBoost classifier in Figure~\ref{fig:confMat}. {Bearing failure entries are classified almost perfectly. The largest fraction of misclassified entries ($>$50\%) is observed for the Horizontal Misalignment state. Also, Shaft Imbalance has 21\% of entries misclassified. Interestingly, all Normal state entries are classified correctly, however the Normal state is often confused with Shaft Imbalance and Horizontal Misalignment.} The potential underlying reasons for such results are detailed in the \nameref{sec:discussion} section.

\subsection{Statistical evaluation}

In order to better understand the results in terms of the experiment comparison as well as address the research questions, we perform the statistical tests and compute the effect sizes. We report these results in Table~\ref{tab:statEval}. In the tests, we expect the absolute errors to be statistically smaller in the alternative hypotheses. 
Median sample values are skewed with respect to the mean values, indicating a non-normal distribution of the data. This is especially evident for logistic regression, RQ1 experiment, where the alternative hypothesis mean error is less than this of the null hypothesis (medians have the opposite relationship), however, the p-value is still 1.0. This is due to the Wilcoxon test nature -- it assesses the number of entries satisfying the condition, rather than the mean. The opposite situation with the medians indicates that even though by the mean absolute error the alternative hypothesis shows supremacy, it is not supported by the majority of entries, and the test confirms that.

Taking the Bonferroni corrections into account, we obtain the corrected significance level $\alpha=0.05/4=0.0125$.

\begin{table}[H]
\small
\caption{The table communicates the statistical test outcomes as well as supporting sample statistics. "Alt." and "Null" correspond to the alternative and null hypothesis, respectively. Below those, we indicate the particular experiment representing the hypotheses. The tests and the sample statistics are computed on the absolute per-entry-per-class classification errors.\label{tab:statEval}}
\begin{tabular}{c|c|c||c|c}
    {Statistics} & \multicolumn{4}{c}{\textbf{Research Question}}\\
    \midrule

   & \multicolumn{4}{c}{\textbf{CatBoost}} \\

   & \multicolumn{2}{c}{{RQ1}} &  \multicolumn{2}{c}{{RQ2}}  \\   
   \midrule
  
       One-tailed Wilcoxon test p-value & \multicolumn{2}{c||}{$<.001$} &  \multicolumn{2}{c}{$<.001$} \\
       Test Statistics & \multicolumn{2}{c||}{245529.0}  & \multicolumn{2}{c}{224899.0} \\
       Effect Size (Hedges \textit{g$_{av}$}) & \multicolumn{2}{c||}{$0.165\pm0.057$}  & \multicolumn{2}{c}{$0.64\pm0.060$} \\
       \midrule
       {} & \multicolumn{4}{c}{\textbf{Test Groups}} \\
         & Alt. & Null &  Alt. & Null \\
         & 50kHz, 5s & 1kHz, 5s &  50kHz, 5s & 50kHz, 0.1s \\
         \midrule
Mean (Absolute Error)  & 4.1$\times10^{-3}$ & 8.4$\times10^{-3}$&  4.1$\times10^{-3}$ &  0.044  \\
Median (Absolute Error)  & 3.5$\times10^{-4}$ & 1.36$\times10^{-3}$& 3.5$\times10^{-4}$ & 1.22$\times10^{-3}$ \\
Standard Deviation & 0.025 & 0.027  & 0.025 & 0.085 \\
    \bottomrule
    \bottomrule
    \end{tabular}

\begin{tabular}{c|c|c||c|c}

   & \multicolumn{4}{c}{\textbf{Logistic Regression}} \\

   & \multicolumn{2}{c}{{RQ1}} &  \multicolumn{2}{c}{{RQ2}}  \\   
   \midrule
  
       One-tailed Wilcoxon test p-value & \multicolumn{2}{c||}{$1.0$} &  \multicolumn{2}{c}{$<.001$} \\
       Test Statistics & \multicolumn{2}{c||}{1050300.0}  & \multicolumn{2}{c}{362824.0} \\
       Effect Size (Hedges \textit{g$_{av}$}) & \multicolumn{2}{c||}{$0.107\pm0.057$}  & \multicolumn{2}{c}{$0.48\pm0.060$} \\
       \midrule
       {} & \multicolumn{4}{c}{\textbf{Test Groups}} \\
         & Alt. & Null &  Alt. & Null \\
         & 50kHz, 5s & 1kHz, 5s &  50kHz, 5s & 50kHz, 0.1s \\
         \midrule
Mean (Absolute Error) & 1.17$\times10^{-3}$ & 3.3$\times10^{-3}$ &  1.17$\times10^{-3}$ &  0.028  \\
Median (Absolute Error) & 7.9$\times10^{-6}$ & 4.0$\times10^{-6}$ & 7.9$\times10^{-6}$ & 1.16$\times10^{-4}$ \\
Standard Deviation & 0.0123 & 0.026  & 0.0123 & 0.077 \\
    \end{tabular}

\end{table}

\section{Discussion}
\label{sec:discussion}
In the current section, we reflect on the obtained results of the experiments. First, we address generally interesting points in the results, then assess the significance of the results, limitations of the study, implications for practitioners, and, finally, discuss the future work. 

In Table~\ref{tab:sigProcHyperOptRes} one notices that EMD shows the best performance for the 5s window with no constraints on IMFs. At the same time, constraints are introduced for the 0.1s window. The potential reason is that spectral components which make sense for the longer window cannot be used for the shorter one, hence should be filtered out. This is also supported by the fact that EMD is known to perform well at filtering noisy non-linear non-stationary data~\cite{peng2011sparse}. 
We see a different picture in the optimisation results for wavelets transformation. Concretely, the lowest scale (highest frequency) is not filtered out for the 50kHz sampling rate, 5s window and filtered for the rest. This indicates that the highest frequencies are utilised by the model only in the setting of the high sampling rate and the larger time window. 
Nevertheless, in Table~\ref{tab:sigProcOrderRes} one notices that both methods are used simultaneously only in 4 cases out of 9. Their sequence is always preserved and contradicts the sequence proposed in~\cite{Han2019FaultCEEMD}. There might be multiple reasons for that, such as different data and experiment settings, slightly different sets of methods. {Interestingly, the non-linear energy operator led to a substantial increase of AICs and was excluded in all the experiment configurations, indicating that this method is not suitable for the considered dataset or the feature space.}
This result suggests that the sequence of the applied methods as well as the methods themselves are quite problem-specific and should be optimised on a per-configuration basis.   
This highlights the importance of the proposed method, allowing systematic, data-driven optimisation of the signal processing pipeline. 

Considering the confusion matrix (Fig.~\ref{fig:confMat}), there is an evident change in the classification performance between different failure modes. Namely, bearing failures are almost perfectly classified, even in the worst-performing configuration (the one reported). We hypothesise that having more rotating elements, bearing failures have a rich vibration spectrum which allows reliable separation of the bearing failures from the rest of the classes.
Two accelerometers mounted close to different bearings likely help to distinguish the failures between the two bearings.
Performance wise the proposed methodology is on par with the existing works on the dataset~\cite{alzghoul2020usefulness,ali2019influence,marins2018improved}. We highlight that the reported performances vary by around 1\% across papers. Considering the size of the dataset these differences translate into a correct classification of extra 20 entries. 
Considering a lack of a uniform framework for comparing the results across studies, the significance of such improvements is rather questionable. 

\subsection{RQ1 -- Sampling rate impact}
In the first research question, we study how the sampling rate affects the model performance (mean absolute error (MAE) of the output per-class probabilities) by comparing the 1kHz and 50kHz sampling rate experiment configurations. We report the p-values in Table~\ref{tab:statEval}. We reject the null hypothesis for the CatBoost estimator, but we cannot reject it for the Logistic Regression case. Hence, we claim that for the CatBoost classifier the higher sampling rate leads to a significantly better performing and more confident model for the considered machine operation frequency range. 
Being significant, effect sizes indicate that these results are likely to hold in the statistical population represented by different machines and configurations. 

\subsection{RQ2 -- Time window impact}
In the second research question, we investigate the impact of the observation time window on the model performance. Based on the p-values in Table~\ref{tab:statEval}, we reject the null hypothesis. Hence, we positively answer the research question claiming that the 5s time window leads to a significantly better classification performance than the 0.1s window.
Similarly to RQ1, the effect sizes are significant. However, the absolute classification performance obtained for the 0.1s is still high. This result suggests that the use of shorter observation windows should be decided on a case-by-case basis.
Of course, when considering industrial applications, there are external sources of vibrations and the amount of labelled data is limited. 
No doubt, it leads to a generally worse classification performance and both the sampling rate and the observation window length should be chosen with the context in mind.

\subsection{Limitations}
\label{sec:discussion:limitations}
It is important to highlight that the proposed experiment design is one of the many ways of achieving the stated objectives.

There are limitations rooting in the design decisions. Concretely, we first perform the hyperparameter optimisation of the signal processing methods, then optimise the order. It might lead to a sub-optimal optimisation result as by this design decision we implicitly assume that the optimal hyperparameters identified for a single method, would hold for the sequence of methods as well. While it is computationally not feasible to perform the hyperparameter optimisation simultaneously with the method sequence optimisation, one should keep this limitation in mind. 

{Another limitation arises when assessing the performance of the models using different metrics. As one can see in Table~\ref{tab:classifResult}, accuracy, precision, recall and f1 scores of the CatBoost estimator are worse for 50kHz, 5s window than for the 1kHz, 5s window, which does not agree with the MAE metric and the outcome of the statistical test in Table~\ref{tab:statEval}. The metrics in Table~\ref{tab:classifResult} consider the output confidence threshold only, hence reflect a different aspect of the model performance than MAE. Since MAE is calculated on the model output class probabilities, it computes how far on average the predicted probabilities are from the true labels, and, consequently, might be giving a more complete picture than the other metrics. We use this metric in the statistical tests as it allows comparing the data in a per-entry fashion, leading to a higher statistical power of the results.}

We perform the experiments on a dataset with labels indicating particular failure modes. The use of the findings is limited to condition monitoring and can hardly be used for predictive maintenance. Also, all the data in the dataset is obtained from a single machine with different failure modes introduced, limiting the generalisability of the findings to multiple machines.
Nevertheless, being general, the proposed approach to the signal processing pipeline optimisation can be used in the predictive maintenance setting and multiple machines data. 

We use a large set of features in the study. By doing so we aim to improve the generalisability of the results. If one uses spectral features outside of the current feature space, there is a chance that the RQ findings will hold, but it would require additional verification. The same holds for the used machine learning models and the obtained configurations of the signal processing pipeline. 

Finally, when optimising the signal processing pipeline, AIC is computed for the logistic regression model. Even though the signal processing optimisation is performed on training data, performance comparison of the two estimators would be biased towards the logistic regression model. The bias is evident in the reported performance of estimators for the 50KHz sampling rate and 0.1s observation window. Such a difference between the estimators might suggest that the proposed method is rather estimator-specific.

\subsection{Implications for practitioners}
Using the demonstrated approach, practitioners are able to optimise their systems, backing the complexity of the pipelines with the data. As we saw, the optimisation allows to effectively detect components which do not contribute to the performance improvement, hence improve the efficiency of the data processing stack making it more suitable for use at the Edge level without a need of transferring large data volumes. The optimisation is computationally demanding, but should be done only once per problem or a family of problems. 

The answers to the research questions contribute to the development of lower-latency and more cost-effective failure classification systems. 
Namely, decrease of the sampling rate would lead to a performance loss, however on an absolute scale the loss is small and might be acceptable in many settings. 
At the same time, decrease of the observation time window leads to a more pronounced performance and model confidence drops (based on effect sizes), hence it is advised to systematically optimise the observation window size on a case-by-case basis. 
From our findings one might conclude that the use of non-linear estimators is not necessary, however this is to be formally studied in an industrial setting, where external sources of noise are present.
Finally, we hope that the methodology used in the study will encourage more reproducibility and comparability in the field.

\subsection{Future work}
There are several pathways the current study can be extended into. As we previously mentioned, the obtained classification performance will likely decrease in the real-world setting. Hence, it would be interesting to extend the findings to industry data with the external vibration sources acting as noise. Considering the complexity of getting the ground truth for a practical number of entries in an industrial setting, the noise can be artificially generated and injected into the available data. This would allow addressing the generalisation of the findings to the real-world setting.  
Another research direction is finding the thresholds of the sampling frequency and the observation window length, below which the classification performance substantially drops. 

\section{Conclusion}
\label{sec:conclusion}
In the current study we proposed a way of optimising the signal processing approach which is further integrated into the machine learning pipeline.
Moreover, we have answer the first research question (RQ1), by finding out that the increased sampling rate significantly improves the classification performance (for 1 of the 2 classifiers). 
Also, we answer the second research question by discovering that the increase of the analysis time window improves the classification performance. 
The findings of the both research questions are complemented by the insignificant effect sizes. It suggests that the improvements coming from the increased sampling rate and the observation window length are likely to generalise to the statistical population. However, the absolute performance drop is moderate, making the use of the lower sampling rates and shorter observation windows feasible for cost effective low latency applications.
Discussing the findings, we detailed the limitations of the experiment design and the findings. 
We conclude that even though lower sampling rates as well as sub-second observation time windows lead to statistically significant classification performance decrease, it is not supported by the effect sizes and their performance on absolute scale is still very high. Hence, they can be used for the failure mode classification in lower budget and lower latency IoT settings.
Finally, we ensure the reproducibility of the study by providing the code base for repeating and extending the reported experiments~\cite{artur_sokolovsky_zenodo}. 


\bibliographystyle{ieeetr}
\bibliography{manualrefs}  

\end{document}